\documentclass[twocolumn,showpacs,preprintnumbers,prx,fleqn,superscriptaddress]{revtex4-2}
\usepackage[english]{babel}
\usepackage[utf8]{inputenc}
\usepackage{SIunits}
\usepackage{graphicx}
\usepackage{dcolumn}
\usepackage{amsmath}
\usepackage{natbib}
\usepackage{bm}
\usepackage{textcomp}
\usepackage{color}
\usepackage{float}
\usepackage{booktabs}
\usepackage{multirow}
\usepackage{tabularx}
\usepackage{adjustbox}
\usepackage{hyperref}
\usepackage{tikz}
\usepackage{makecell}

\begin{document}

\title{High pressure tuning of magnon-polarons in the layered antiferromagnet FePS$_{3}$}

\author{Amit Pawbake}
\thanks{These authors contributed equally}
\affiliation{LNCMI, UPR 3228, CNRS, EMFL, Université Grenoble Alpes, 38000 Grenoble, France}
\author{Thomas Pelini}
\thanks{These authors contributed equally}
\affiliation{LNCMI, UPR 3228, CNRS, EMFL, Université Grenoble Alpes, 38000 Grenoble, France}
\author{Alex Delhomme}
\affiliation{LNCMI, UPR 3228, CNRS, EMFL, Université Grenoble Alpes, 38000 Grenoble, France}
\author{Davide Romanin}
\affiliation{Sorbonne Université, CNRS, Institut des Nanosciences de Paris, UMR 7588, F-75252 Paris, France}
\author{Diana Vaclavkova}
\affiliation{LNCMI, UPR 3228, CNRS, EMFL, Université Grenoble Alpes, 38000 Grenoble, France}
\author{Gerard Martinez}
\affiliation{LNCMI, UPR 3228, CNRS, EMFL, Université Grenoble Alpes, 38000 Grenoble, France}
\author{Matteo Calandra}
\affiliation{Sorbonne Université, CNRS, Institut des Nanosciences de Paris, UMR 7588, F-75252 Paris, France}
\affiliation{Dipartimento di Fisica, Università di Trento, Via Sommarive 14, 38123 Povo, Italy}
\author{Marie-Aude Measson}
\affiliation{Institut Neel, Université Grenoble Alpes, 38000 Grenoble, France}
\author{Martin Veis}
\affiliation{Faculty of Mathematics and Physics, Institute of Physics, Charles University, Ke Karlovu 5, 121 16 Prague 2, Czech Republic}
\author{Marek Potemski}
\affiliation{LNCMI, UPR 3228, CNRS, EMFL, Université Grenoble Alpes, 38000 Grenoble, France}
\author{Milan Orlita}
\affiliation{LNCMI, UPR 3228, CNRS, EMFL, Université Grenoble Alpes, 38000 Grenoble, France}
\author{Clement Faugeras}
\email{clement.faugeras@lncmi.cnrs.fr} \affiliation{LNCMI, UPR 3228, CNRS, EMFL, Université Grenoble Alpes, 38000 Grenoble, France}

\date{\today }

\begin{abstract}
 Magnetic layered materials have emerged recently as promising systems to introduce magnetism in structures based on two-dimensional (2D) materials and to investigate exotic magnetic ground states in the 2D limit. In this work, we apply high hydrostatic pressures up to $P~\sim8.7$~GPa to the bulk layered antiferromagnet FePS$_3$ to tune the collective lattice excitations (phonons) in resonance with magnetic excitations (magnons).
Close to $P=4$~GPa, the magnon-phonon resonance is achieved and the strong coupling between these collective modes leads to the formation of new quasi-particles, the magnon-polarons, evidenced in our low temperature Raman scattering experiments by a particular avoided crossing behavior between the phonon and the doubly degenerate antiferromagnetic magnon. At the pressure-induced magnon-phonon resonance, three distinct coupled modes emerge. As it is mainly defined by intralayer properties, we show that the energy of the magnon is nearly pressure independent. We additionally apply high magnetic fields up to $B=30$~T to fully identify and characterize the magnon excitations, and to explore the different magnon-polaron regimes for which the phonon has an energy lower-, equal to-, or higher- than the magnon energy. The description of our experimental data requires introducing a phonon-phonon coupling not taken into account in actual calculations.
\end{abstract}

\maketitle

\section{Introduction}
Quantum mechanics describes how new quasi-particles can emerge from the hybridization of existing particles through a strong coupling mechanism that mixes the initial (bare) states. Depending on the initial energy detuning between the two bare modes, the newly formed quasi-particles exhibit either the character of the individual entities, or a mixed character of the two excitations. Strong coupling between collective excitations appears as a way to create and manipulate new quasi-particles with distinct properties. A direct way to reveal a strong coupling is to tune the selected modes (quasi-particles) into resonance. The coupling is then manifested in the characteristic avoided crossing behavior between the coupled modes at resonance. This methodology was successfully applied to plasmons and phonons~\cite{Mooradian1966, Wysmolek2006}, to excitons and optical modes of microcavities forming exciton-polaritons~\cite{Weisbuch1992}, to electrons and phonons leading to polarons~\cite{Horst1983}, or magnons and phonons leading to magnon-polarons~\cite{Hagi1999}. When one of the two modes is doubly degenerate with its individual components non-interacting with each other, the energy spectrum of the coupled system is modified in a very peculiar way, leaving one of the degenerate components unaltered in energy but its character changes as a function of the energy detuning.

Magnon-polarons, coupled magnon and phonon excitations, are of prime importance for future spintronic applications, and in particular those based on antiferromagnetic materials~\cite{Baltz2018,Rezende2019}, as they result from the spin-lattice interaction, and are central to describe the properties of spin waves in magnetically ordered solids~\cite{Kikkawa2016,Benedetta2017,Hayashi2018}. Because the magnon and phonon energies at $\textbf{k}=0$, where $\textbf{k}$ is the momentum of the excitation, are most often different, magnon-polarons are evidenced at $\textbf{k}\neq0$ by neutron scattering techniques~\cite{Haoran17,Sukhanov19} or using patterned surfaces and diffraction effects~\cite{Godejohann2020}. Zone-center magnon-polarons are evidenced by applying a magnetic field to change the magnon energy and to bring it in coincidence with a phonon excitation~\cite{Allen1969} while probing both phonon and magnon excitation spectra by optical means. In the case of FePS$_3$, a layered antiferromagnet, a series of magnon-polarons involving the three phonon modes with energy lower than the magnon excitation, have been recently observed in high magnetic fields~\cite{Liu2021,Vaclavkova2021,zhang2021}.

High pressure environments are particularly relevant for investigations of layered materials as they provide a mean to tune the interlayer distance and all interactions that directly depend on the magnitude of the van der Waals gap~\cite{Xia2020,Zhao2021,Ma2021,Yankowitz2018,Yankowitz2019}. Applying high pressure to a material also induces a modification of the phonon spectrum by directly affecting the bond lengths. Recent X-ray diffraction investigations of the pressure evolution of the lattice parameters of FePS$_3$~\cite{Haines2018,Wang2018} or investigations of the phonon spectrum of a heterobilayer of two dimensional materials~\cite{Xia2020}, suggest that applying hydrostatic pressure to a layered material mainly changes the interlayer distance, leaving the in-plane degrees of freedom weakly altered. In a first approximation, applying hydrostatic pressure to a layered material mainly reduces its inter-layer spacing, the van der Waals gap, while leaving intralayer parameters unchanged.

FePS$_3$ is a layered phosphorous trichalcogenide from the broad family of MPX$_3$ compounds (M = Fe, Mn, Ni, Co and X = S, Se). It is an antiferromagnet with a N\'{e}el temperature $T_{N} = 118$~K~\cite{LeFlem1982}. It crystalizes in the monoclinic system with C2/m space group (see Fig.~\ref{Fig1}a) and, because of the strong single Fe$^{2+}$ ion anisotropy, is considered as an Ising-type zig-zag antiferromagnet with magnetic moments perpendicular to the layer plane~\cite{Lancon2016,Wildes20a}. Its magnetic properties persist down to the monolayer limit~\cite{Wang2016} with an increase of $T_N$ for the monolayer. At $\textbf{k}=0$, magnons in FePS$_3$ have an unusually high energy of E$_M \sim 122$~cm$^{-1}$~\cite{Sekin1990,McCreary2020}. Bulk FePS$_3$ shows two structural phase transitions under pressure~\cite{Wang2016,Haines2018}: a first one close to $P\sim4$~GPa where the system transforms into a second C2/m phase named HP-I, and another one above $P\sim14$~GPa labelled HP-II in which bulk FePS$_3$ transforms to a metallic trigonal P31m phase. These phases have recently been explored by Raman scattering at room temperature~\cite{Sood2022} and the critical pressure corresponding to the first transition has been determined to be $P=4.6$~GPa. The magnetic properties of these phases have been described by neutron scattering under high pressure~\cite{Ko83,Rule2007,Coak2021}. The ground state of bulk FePS$_3$ at ambient pressure is composed of ferromagnetic zig-zag chains of Fe atoms along the a axis, which are antiferromagnetically coupled both in-plane and with the adjacent layers, see Fig.~\ref{Fig1}a. When applying hydrostatic pressure, the intralayer antiferromagnetic order persists but the interlayer interaction becomes ferromagnetic in HP-I phase. For pressures above $14$~GPa, a metallic phase with short-range magnetic order is observed.

Under hydrostatic pressure, the magnon excitation M of FePS$_3$, which can be identified from its coupling to an external out-of-plane magnetic field, has mainly an intralayer character~\cite{Lancon2016, wildes20b} and its energy depends only very weakly on the applied pressure. In contrast, the energies of all phonon modes increase when applying pressure. In this work, we apply high pressure to bulk FePS$_3$ in its low temperature antiferromagnetic phase and, by using magneto-Raman scattering spectroscopy~\cite{Fleury1968} techniques, we reveal a strong coupling between magnon and phonon excitations by tuning the phonon energy in resonance with the magnon. These results are discussed in the frame of a simplified magnon-phonon interaction model presented in details in Ref.~\cite{Vaclavkova2021}, including the two degenerate magnon excitations and the three low energy optical phonons P$_{1-3}$ at $\textbf{k}=0$. One of the consequences of the magnon-phonon interaction is the lifting of the magnon degeneracy. This effect, observable at ambient pressure~\cite{Vaclavkova2021}, is strongly enhanced when the P$_3$ phonon is tuned to resonance with the magnon by increasing pressure. The energies of the three modes correspond to an avoided crossing typical of the strong coupling regime, with one of the coupled modes remaining at the bare magnon energy, unaffected by the coupling. Applying a magnetic field allows the unambiguous identification of the magnon excitation when pressure is applied and allows for the observation of magnon-polarons in the different situations where the phonon energy is smaller, equal to, or higher than that of the magnon, within the same material. These experiments enable the determination of the magnon-phonon coupling constant and for their evolution with hydrostatic pressure.

\section{Methods}

\subsection{Experimental}

\begin{figure}
\centering
\includegraphics[width=0.95\linewidth,angle=0,clip]{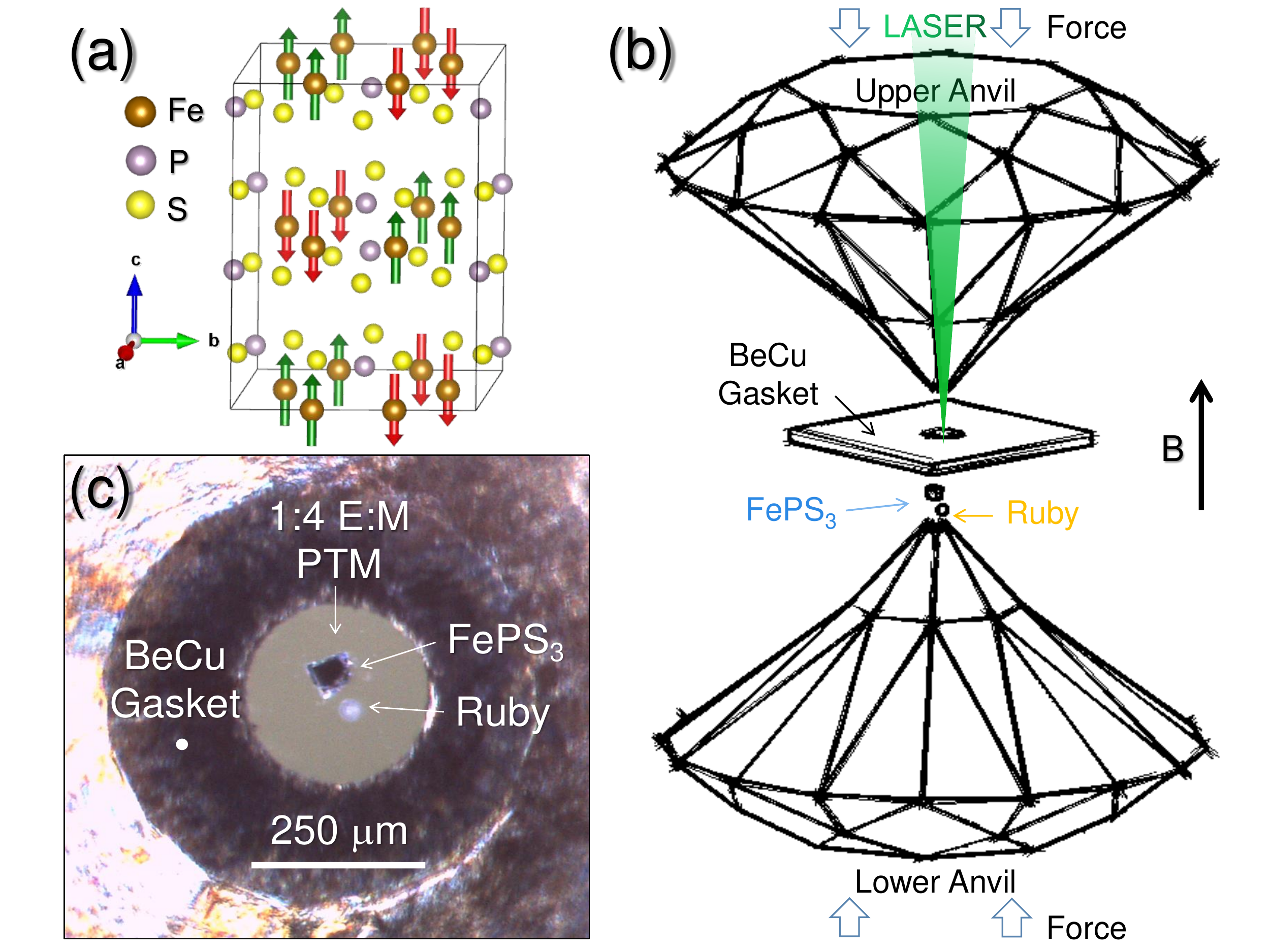}
\caption{a) Crystal lattice of FePS$_3$ with the magnetic moments of the two zig-zag chains in green and red. b) Schematics of the experiment showing the diamonds of the diamond anvil cell, the FePS$_3$ sample and the ruby ball marker. c) Optical photograph of the pressure chamber showing the metallic gasket with a hole of diameter 250$\mu m$.
\label{Fig1}}
\end{figure}

A flake of commercial bulk FePS$_3$ (HQ graphene) together with a ruby crystal used as a pressure marker~\cite{Shen2020}, have been inserted in the pressure chamber of a diamond anvil cell (DAC) filled with a solution of methanol:ethanol (4:1), see Fig.~\ref{Fig1}b,c). The DAC is placed on piezo motors allowing for spatial displacements in three directions with sub-micrometer resolution. For magneto-optical measurements, we use a home-made optical experimental setup which detailed description can be found in Ref.~\cite{Breslavetz2021}. It is based on free-beam optics with a $12$~mm working distance objective (numerical aperture NA$=0.35$) to excite the sample and to collect the scattered signals. With the piezo motors, we can move the DAC below the excitation laser and spatially map the optical response of the pressure chamber or investigate the properties of a well defined location, with a micrometer spatial resolution. The setup is placed in a closed metallic tube filled with $\sim150$~mbars of helium exchange gas at room temperature. The system is coupled to a $70$~cm focal grating spectrometer equipped with a nitrogen cooled charge coupled device (CCD) camera. Three volume Bragg filters are used in transmission before the spectrometer in order to reject stray light. Experiments are performed at liquid helium temperature with a $\lambda=515$~nm excitation from a fiber-based laser keeping the optical power below $1$~mW. For magneto-Raman scattering experiments, the experimental set-up is placed in the $50$~mm diameter bore of a $20$~MW resistive magnet at LNCMI-CNRS producing fields up to $B=31$~T, or that of a $14$~T superconducting solenoid. Typical acquisition times with our equipment are of the order of $1$ to $10$ minutes per spectrum.

\subsection{Theory}

Spin polarized first principles density functional theory (DFT) calculations have been carried out by using the PBE0 hybrid exchange-correlation functional and the CRYSTAL~\cite{DovesiQM2014,DovesiW2018} software. We used the triple-polarized Gaussian type basis set~\cite{OliveiraCC2019} with real space integration tolerances of 8-8-8-8-16 and an energy tolerance of $10^{-10}$ Ha for the total energy convergence. The Brillouin zone was sampled with a uniform grid of k-points of $8\times8\times8$. Dispersive Van der Waals interactions were treated in the Grimme-D3 parametrization~\cite{GrimmeJCP2010}.

\section{Results and discussion}

\subsection{Pressure induced magnon-polarons}

Magnon-polarons have recently been evidenced in bulk FePS$_3$ by magneto-Raman scattering and infrared spectroscopy techniques~\cite{Liu2021,Vaclavkova2021,zhang2021}. The strong coupling between magnons and phonons affects the energies of both excitations for any energy detuning, and this effect is strongly enhanced when the energies of both excitations are resonant. The application of a magnetic field adds a Zeeman contribution to the magnon energy that allows tuning its energy. The antiferromagnetic magnon double degeneracy is lifted and one of the two magnon-branches can be tuned in resonance with a phonon. To investigate the physical case where a doubly degenerated magnon resonantly interacts with a phonon, one has to find a way to tune the phonon spectrum independently of the magnon excitation. We show that applying high hydrostatic pressure allows to tune directly the resonance of phonon and magnon.

The bare magnon energy can be calculated from a Heisenberg Hamiltonian with a large single ion anisotropy~\cite{Lancon2016}. The magnon energy is of $122$~cm$^{-1}$~\cite{Sekin1990,McCreary2020,Liu2021,Vaclavkova2021}, defined, within this model, by the three in-plane exchange parameters $J_i$, where $i=1..3$, which represent the in-plane first-, second- and third- nearest neighbors exchange parameters by one interplanar exchange parameter $J'$, and by the single ion anisotropy D. Because $J'$ is small with respect to in-plane exchange parameters, magnons in FePS$_3$ can be seen, in a first approximation, as intralayer excitations. In contrast, phonons are mostly delocalized over many layers and applying hydrostatic pressure mainly affects the van der Waals gap that changes significantly the phonon spectrum while leaving magnon excitations unaffected.

\begin{figure}
\centering
\includegraphics[width=0.95\linewidth,angle=0,clip]{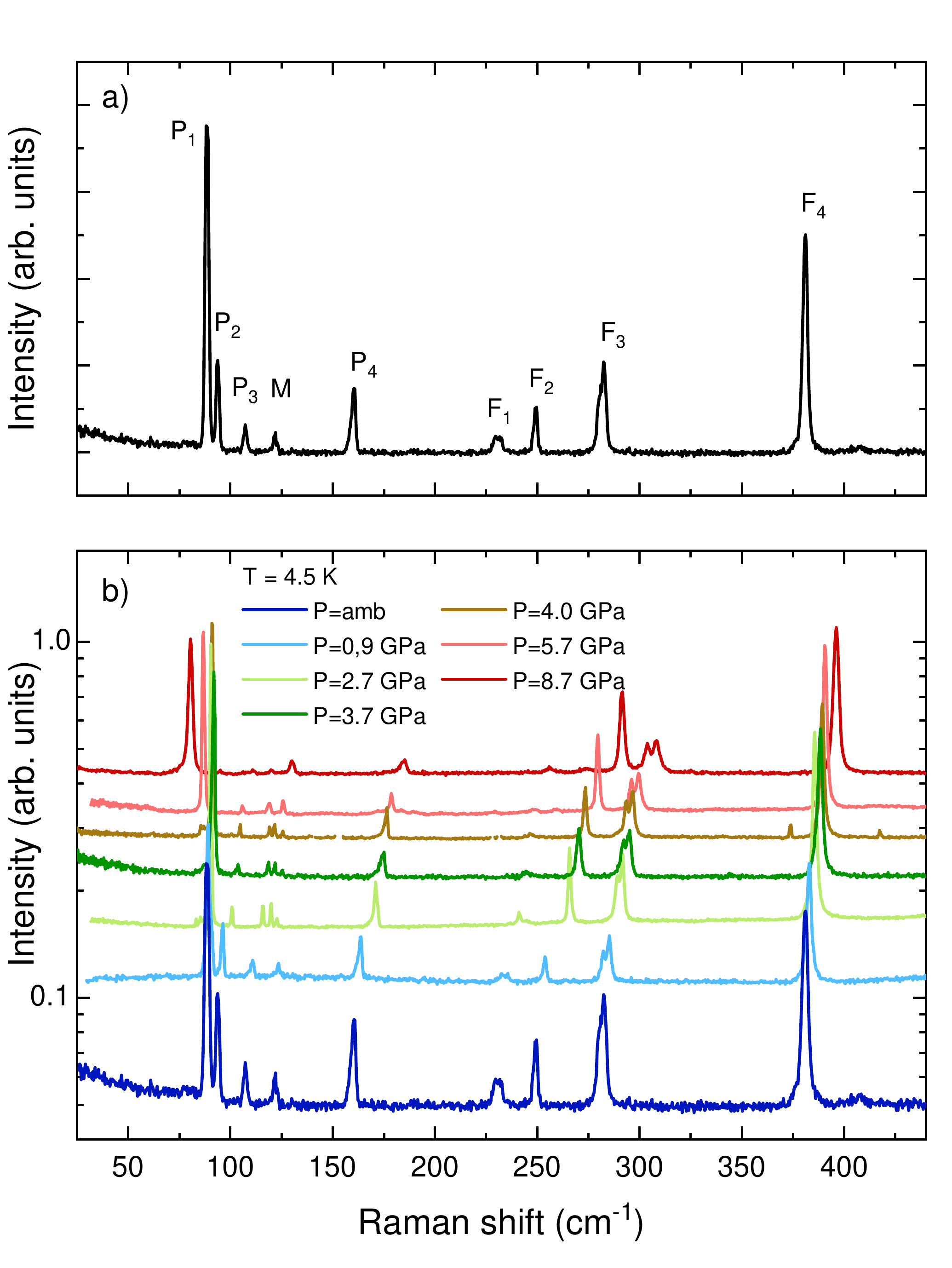}
\caption{a) Low temperature ($T=4.5$~K) Raman scattering spectrum of bulk FePS$_3$ measured at ambient pressure. The different phonon modes are labeled P$_i$ with $i=1-4$ and F$_j$ with $j=1-4$ and the doubly degenerated antiferromagnetic magnon excitation is labeled M. b) Low temperature Raman scattering spectra measured at different hydrostatic pressures up to $P=8.7$~GPa with logarithmic scaling. Spectra are shifted for clarity.
\label{Fig2}}
\end{figure}

The low temperature Raman scattering response of bulk FePS$_3$ in its antiferromagnetic phase at ambient pressure is depicted in Fig.~\ref{Fig2}a. It includes different phonon contributions P$_{1-4}$, F$_{1-4}$ and a magnon excitation labelled M. The high energy F$_i$ phonons are mostly related to vibrations of the $(P_2S_6)^{4-}$ groups~\cite{Hashemi2017}. The three Raman peaks P$_1$, P$_2$, P$_3$ and M below $140$~cm$^{-1}$ have been extensively described~\cite{Scagliotti1987,Sekin1990,Lee2016,Wang2016}. P$_{1-3}$ are phonon modes which involve the movement of Fe$^{2+}$ ions and are hence related to the magnetic properties of FePS$_3$. They become Raman active below the N\'{e}el temperature most probably because the low temperature magnetic order implies an additional periodicity, a doubling of the unit cell with four inequivalent magnetic sublattices, which causes the folding of the phonon Brillouin zone. Acoustic phonons from high symmetry points of the Brillouin zone are then folded on the $\Gamma$ point and become Raman active~\cite{Lee2016,Kargar2020,Liu2021}. The magnon excitation M has been identified by its particular temperature evolution~\cite{Sekin1990,Lee2016} or when applying an out-of-plane magnetic field~\cite{McCreary2020}. In the latter case, the two fold degeneracy of the antiferromagnetic magnon is lifted and two magnon branches appear, splited by $2 g \mu_B B$ where $\mu_B$ is the Bohr magneton and $B$ the magnetic field, with a $g$-factor close to $g=2.1$~\cite{McCreary2020,Liu2021,Vaclavkova2021,zhang2021}. For high enough magnetic fields ($B >12$~T), the low-energy magnon branch in bulk FePS$_3$ reaches the energy of P$_{1-3}$ optical phonons and these collective excitations hybridize, leading to the formation of magnon-polarons. The experimental signature of these new quasi-particles is a strong avoided crossing of the two modes when the low energy magnon branch is tuned to the phonon energies~\cite{Liu2021,Vaclavkova2021}.

The evolution of the low-temperature Raman scattering response of bulk FePS$_3$ when increasing pressure is presented in Fig.~\ref{Fig2}b. For pressure values below $P=3$~GPa, the energies of the P and F phonon peaks increase linearly (see Fig.~\ref{Fig3}b and Fig.~S1). The corresponding slopes are indicated in Table.~S1 of the supplementary material. The magnon excitation can be identified at any applied pressure by imposing a magnetic field while measuring the Raman scattering response. The energy of the magnon excitation M remains rather constant at $\sim 122$~cm$^{-1}$ up to $P=4$~GPa. At higher pressures, it downshifts by about $2$~cm$^{-1}$. P$_1$ and P$_2$ phonon modes evolve with different slopes which makes their energy separation increasing with increasing pressure. Simultaneously, the intensity of P$_2$ decreases significantly.

The low temperature phonon band structure is still not well known as the magnetic order induces a unit cell which contains 20 atoms (Niggli cell). In order to clarify the origin of the P$_{1-3}$ phonons, we have carried out ab-initio density functional theory (DFT) calculations at zero and finite pressure in the 0-2 GPa range. In Table~\ref{table:1} we report the frequencies for the four low energy phonon modes at ambient pressure compared to the experimental values.  The normal modes normalized to classical amplitudes of each mode are reported in Fig.~\ref{Fig4}.

The first two vibrational modes (normally labeled P$_1$ and P$_2$) show a predominant out-of-plane displacement of Fe atoms with the sole difference linked to the in-plane and out-of-plane motion of S and P atoms for P$_1$ and P$_2$ respectively (Fig.~\ref{Fig4} P$_1$ - $B^1_g$ and P$_2$ - $B^2_g$). On the other hand, P$_4$ does not show any contribution coming from Fe atoms, leaving P and S atoms in-plane and out-of-plane displacements only (Fig.~\ref{Fig4} P$_4$ - $A^2_g$). Finally, the assignment of the Raman active P$_3$ mode is controversial in the literature due to the magnetic order and the reduced symmetry of the crystal. As a matter of fact,  the P$_3$ mode has been attributed either to an out-of-plane~\cite{Liu2021} or to an in-plane~\cite{zhang2021} displacement pattern of the Fe atoms. In the energy region of the P$_3$ mode, there are two Raman active modes close to the experimental value of $\sim107$ cm$^{-1}$, see Table.~\ref{table:1}. The first one, P$_3$ - $A^1_g$, involves in-plane Fe motion while the second one, P$_3$ - $B^3_g$, shows an out-of-plane displacement (Fig.~\ref{Fig4}). While the in-plane mode is the closest to the experimentally measured value at ambient pressure, the out-of-plane one shows an evolution as a function of pressure closer to the experimental data (see Table S2 of the supplementary material). The controversial results found in the literature are probably due to the extreme sensitivity of the phonon frequency to the exchange correlation functional but also to the use, in some cases, of a simplified structure composed of a single layer only.


\begin{figure}
\centering
\includegraphics[width=0.9\linewidth,angle=0,clip]{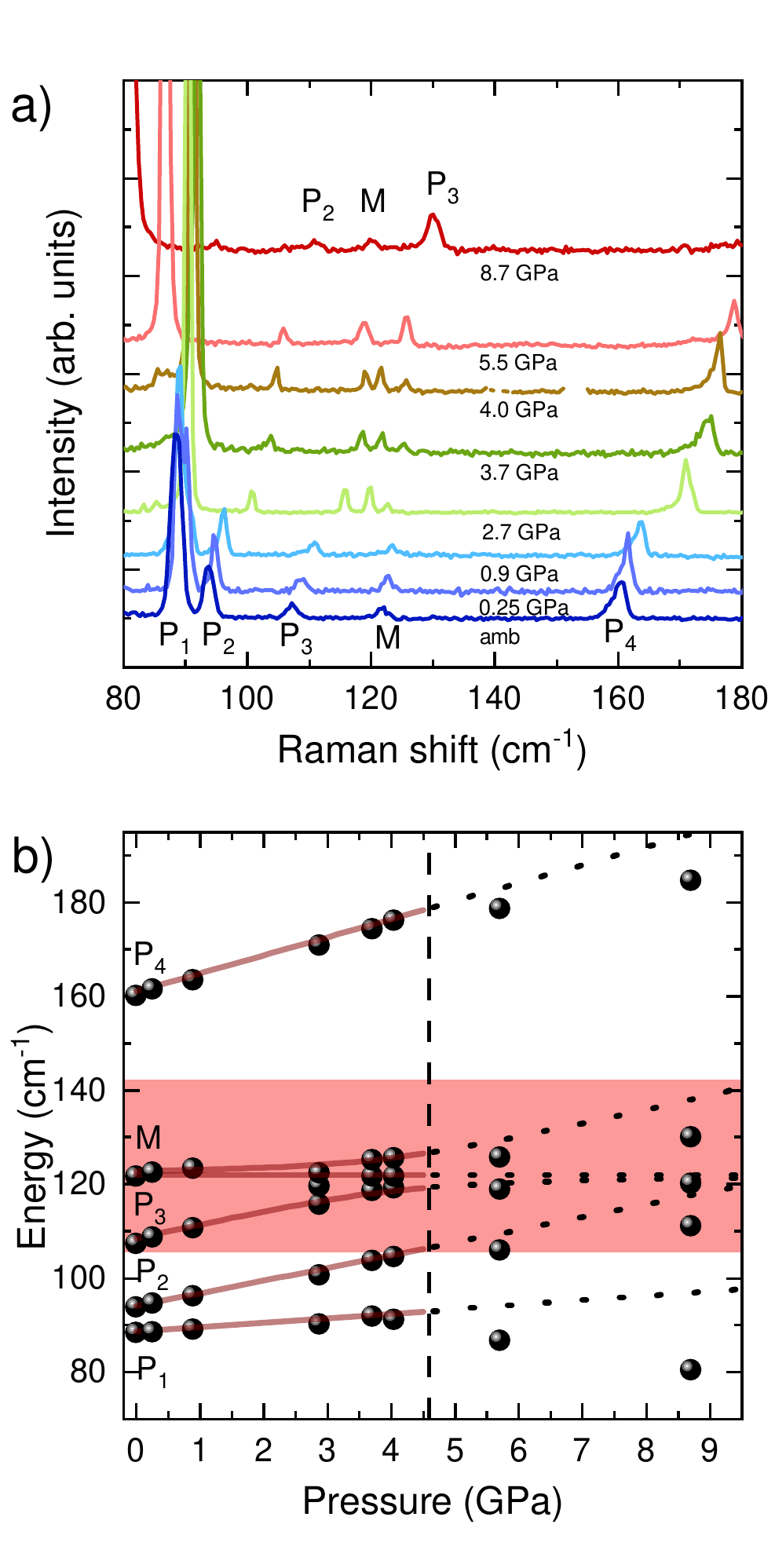}
\caption{a) Same spectra as the one presented in Fig.~\ref{Fig2}a) but focusing on the low energy part of the spectrum. b) Evolution of the maxima of the peaks of the Raman scattering response as a function of the applied pressure. Solid lines are the results of a calculation using a coupling constant $\lambda_3=2.8\pm0.3$~cm$^{-1}$. The vertical dashed line indicates P$=4.6$~GPa, the pressure induced structural phase transition. The horizontal dashed lines are the solutions of our model beyond the phase transition at $P=4.6$~GPa, which is not considered in the calculations. The colored region highlights the region where the triple crossing occurs.
\label{Fig3}}
\end{figure}

\begin{table}[h]
    \centering
\begin{tabular}{ c c c c}
\hline
\hline
Mode & Experimental (cm$^{-1}$)  & Theory (cm$^{-1}$) & Symmetry\\
\hline
P$_1$ & 88.4 & 94.63 & $B^1_g$\\
P$_2$ & 93.8 & 94.70 & $B^2_g$\\
\multirow{2}{*}{P$_3$} & \multirow{2}{*}{107.3} & 115.44 & $A^1_g$\\
                                                                  &    & 125.36 & $B^3_g$\\
P$_4$ & 160.1 & 159.67 & $A^2_g$\\
\hline
\hline
\end{tabular}
\caption{Experimental and theoretical (DFT) frequencies at $P=0$ GPa. For the P$_3$ mode we report the two Raman active phonon modes
closer to the experimental Raman frequency}
\label{table:1}
\end{table}

\begin{figure}[h]
\centering
\includegraphics[width=0.95\linewidth,angle=0,clip]{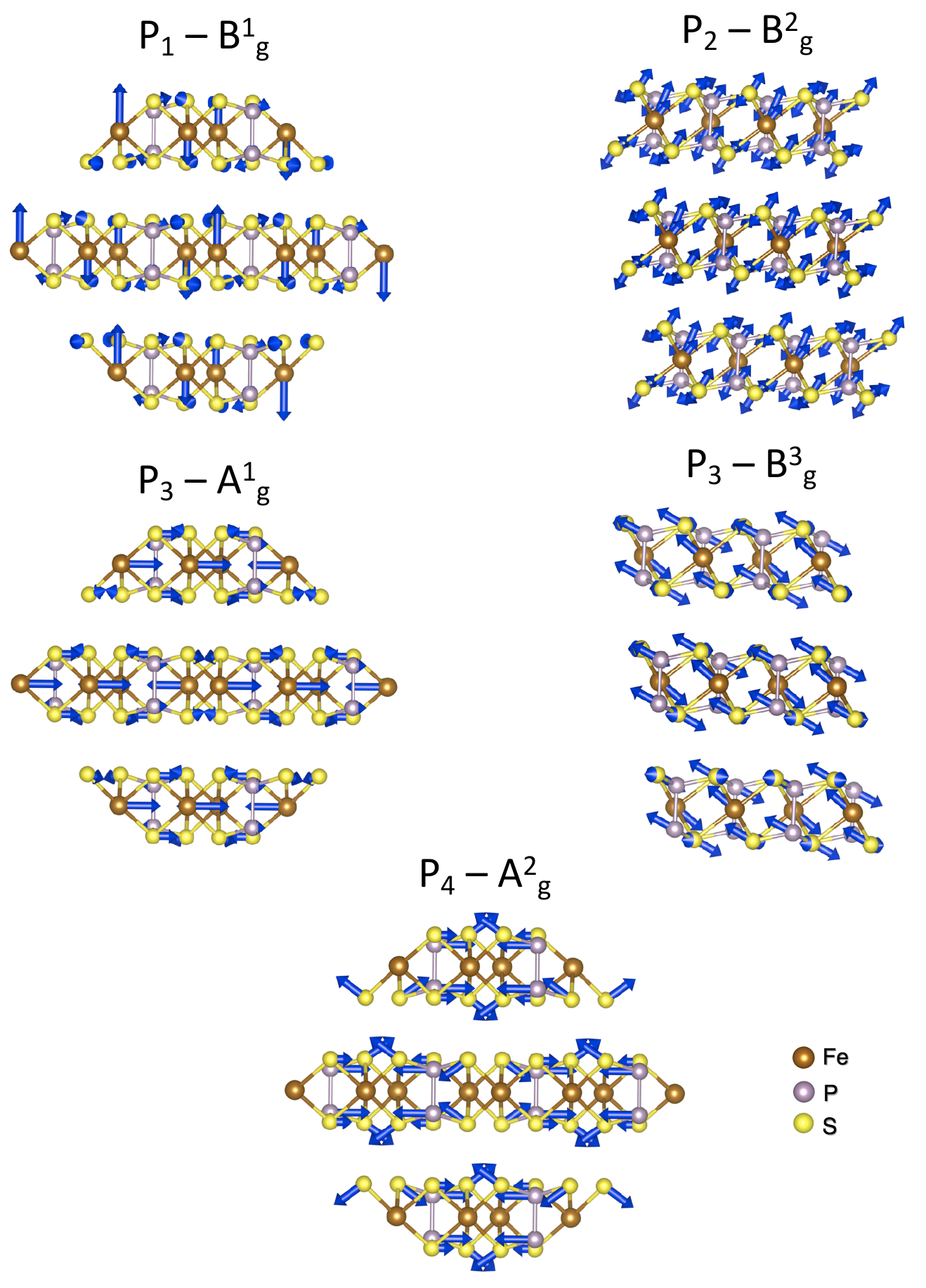}
\caption{Normal modes normalized to classical amplitudes corresponding to the four low energy phonon modes calculated via DFT and PBE0 exchange-correlation functional. The brown balls are Fe atoms, the purple balls are P atoms while the yellow balls are S atoms. The blue arrows represent the direction and amplitude of the displacement of the atom it is attached to.
\label{Fig4}}
\end{figure}

From our Raman scattering data, we deduce that the P$_3$ phonon energy increases with a rate close to $3$~cm$^{-1}$/GPa and reaches the magnon energy for pressures in the range of $3-4$~GPa. At resonance, the Raman scattering spectrum changes drastically with the appearance of a third mode, as it is demonstrated in Fig.~\ref{Fig3}a. For higher pressures, the Raman scattering spectrum recovers its initial shape with a single peak for the magnon, and a single peak for the P$_3$ phonon, at energies then higher than the magnon energy. We interpret this profound change of the Raman scattering spectrum as the spectroscopic signature of $\Gamma$-point magnon-polaron resonance, observed here in the case of a doubly degenerate antiferromagnetic magnon-excitation as no magnetic field is required to achieve the magnon-phonon resonance. The magnon-phonon interaction lifts the magnon double degeneracy and induces a splitting of $\approx 7$~cm$^{-1}$ at the resonance.  This observation confirms the magnon-phonon interaction as a possible origin for the $1$~cm$^{-1}$ zero-magnetic field magnon splitting reported in Ref.~\cite{Vaclavkova2021}, which value depends on the strength of the magnon-phonon interaction as well as on the magnon-phonon energy detuning.

To explain these results, we have used the model presented in Ref.~\cite{Vaclavkova2021} using the following Hamiltonian:

\begin{equation}
H_{5\times5} = \begin{bmatrix}
M_+ & 0 & \delta_1 & 0 & \delta_3\\
0 & M_- & \delta_1 & 0 & \delta_3 \\
\delta_1 & \delta_1 & P_1 & \beta & 0 \\
0 & 0 & \beta & P_2 & 0 \\
\delta_3 & \delta_3 & 0 & 0 & P_3
\end{bmatrix}
\label{Ham}
\end{equation}

where $M_{\pm}$ are the energies of the two antiferromagnetic magnon branches, P$_i$ are the phonon energies, $\delta_{1,3}$ are the magnon-phonon coupling parameters for P$_1$ and P$_3$ while $\beta$ describes the coupling of P$_1$ and P$_2$. This model utilizes the Heisenberg Hamiltonian of the magnetic system and dispersionless phonons. Following the Holstein-Primakoff and the Bogolyubov transformations for bosonic operators one can derive the eigenstates of the coupled magnon-phonon modes. We have adapted this phenomenological model by imposing a linear evolution of all phonon modes with their respective experimentally determined rate R.  We have used as starting parameters the magnon-phonon coupling parameters $\lambda_i$ determined from Raman scattering experiments at ambient pressure~\cite{Vaclavkova2021} and have slightly adjusted the coupling parameter $\lambda_3$. The results of this calculation are presented in Fig.~\ref{Fig3}b together with the experimental positions of the peaks of the Raman scattering response. This model describes the resonance profile experimentally observed and brings a value for the coupling parameter of $\lambda_3=2.9\pm0.3$~cm$^{-1}$. This value is in line with the coupling constant deduced from magneto-Raman scattering experiments at ambient pressure~\cite{Vaclavkova2021}. Other magnon-phonon coupling parameters $\lambda_i$, less relevant in this situation because of the large magnon-phonon P$_i$ ($i\neq3$) energy detuning, cannot be evaluated and are considered as unchanged.

At resonance, the effect of the magnon-phonon interaction is maximized, leading to a $7$~cm$^{-1}$ splitting of the two magnon components and to the observation of three distinct peaks, representative of coupled magnon-phonon modes. The energy of the central mode weakly depends on pressure nor on the coupling strength. Following the theoretical model, its energy coincides with the bare magnon energy, and is inherent to coupled multi modes systems with two degenerate non-interacting modes, in the present case, the two components of the magnon. We present in Fig.~S2 of the supplementary material a comparison with a simpler Hamiltonian based on a $3\times3$ matrix. One can also note that despite the difference in the involved energies (structural vs magnetic), this magnon-polaron resonance occurs at a pressure comparable to that of the first structural transition in bulk FePS$_3$~\cite{Haines2018,Sood2022}. For pressures above P$=4.6$~GPa where the structural phase transition is expected, the low-energy phonon spectrum (below $180$~cm$^{-1}$) changes significantly: the energy of P$_1$ decreases and is lower than the one at ambient pressure ($E_1=80$~cm$^{-1}$ at P$=8.7$~GPa), the energies of P$_2$ and of P$_3$ continue increasing but with a lower rate. As it is shown in Fig.~S1 of the supplementary material, the evolution with pressure of the F$_i$ phonon modes remains linear, with no pronounced change of the rate at $P=4.6$~GPa.

Under such conditions, the P$_3$ phonon has an energy larger than the magnon and the magnon energy is reduced by $2$~cm$^{-1}$ (at $P=5.7$ and $8.7$ GPa). When the hydrostatic pressure is increased and the P$_3$ phonon energy approaches the magnon energy, we do not observed any blueshift of the magnon mode. As a result, we understand this softening of the magnon energy as the signature of the magnetic phase transition observed in neutron scattering experiments~\cite{Coak2021}. When applying pressure to bulk FePS$_3$, the van der Waals gap decreases and the interlayer interaction becomes ferromagnetic in the HP-I phase. The values of the interlayer magnetic exchange coupling parameter $J'$ and of the single ion anisotropy $D$ are changed which results in a modification of the magnon energy. This decrease of the magnon energy of $1.6\%$ is the signature of the magnetic HP-I phase. The pressure independent magnon-phonon mode is of particular interest as, from a magnon-like character at ambient pressure, it changes to phonon-like close to the resonant hydrostatic pressure, and then back to a magnon-like character. Meanwhile, its energy remains quasi constant.

\begin{figure}
\centering
\includegraphics[width=0.95\linewidth,angle=0,clip]{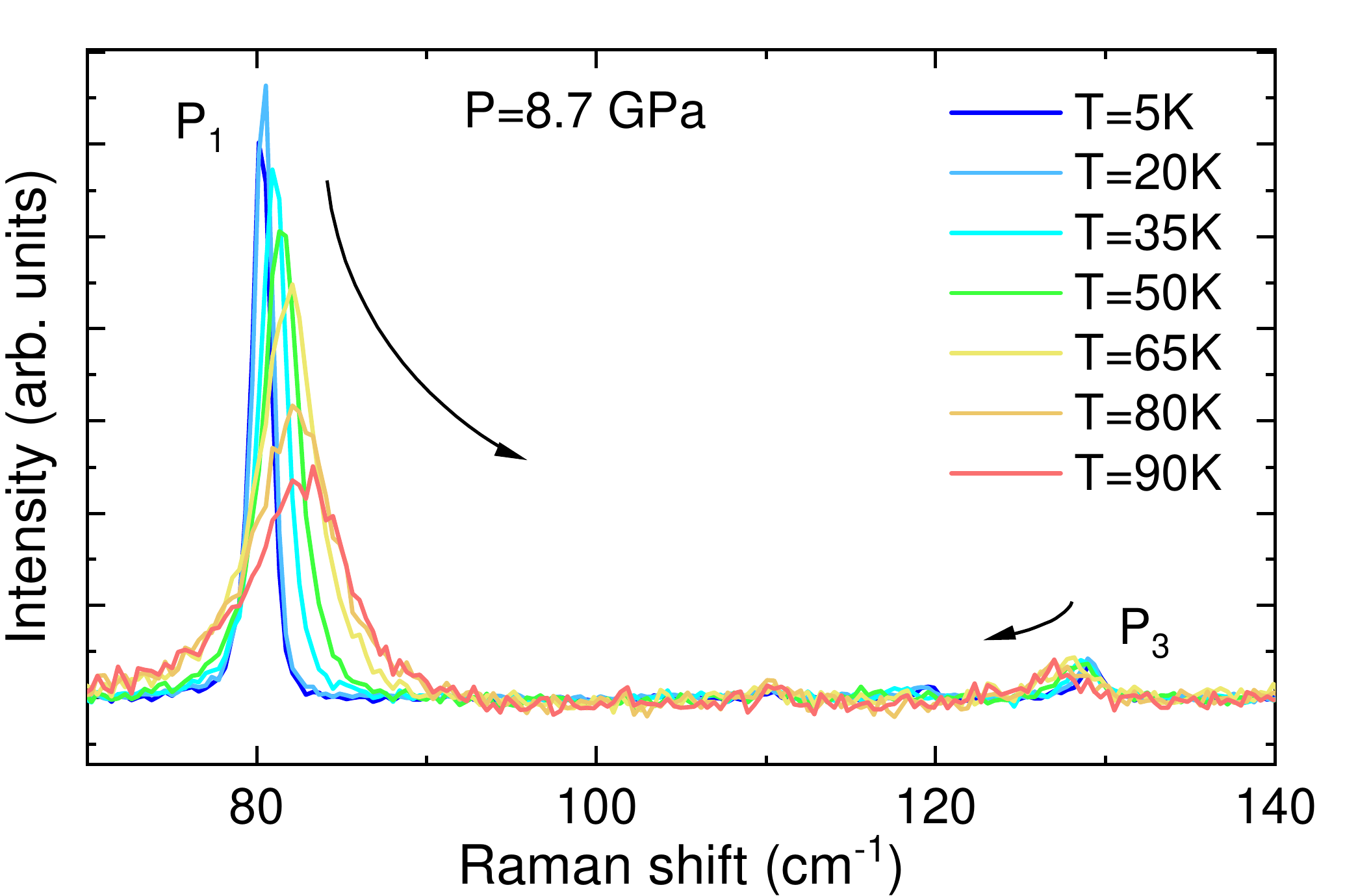}
\caption{Evolution of the Raman scattering response as a function of the temperature at $P=8.7$~GPa. The P$_1$ and P$_3$ phonons are indicated while the acquisition time does not allow for the observation of the weak P$_2$ and M features. The arrows are guides for the eyes.
\label{Figtemp}}
\end{figure}

The low temperature phonon spectrum observed at $P=8.7$~GPa is representative of the HP-I high pressure phase described in Ref.~\cite{Haines2018}. Similar to the ambient pressure situation~\cite{Sekin1990,Lee2016}, the P$_1$ Raman scattering peak observed at $E_1=$80~cm$^{-1}$ is influenced by the magnetic order in bulk FePS$_3$. This can be deduced from its evolution when rising temperature, see Fig.~\ref{Figtemp}. Both $P_1$ and $P_3$ variations show a characteristic coupling to the magnon which shifts their energy away from $E_M$ by $2$~cm$^{-1}$ in the investigated range of temperature. This temperature dependence is a strong indication that, as it has been observed at ambient pressure, the low energy Raman scattering features are intimately related to the magnetic interaction in the solid.

In this section, we have demonstrated that the application of high hydrostatic pressures on bulk FePS$_3$ modifies the phonon excitation spectrum while leaving the magnon excitation nearly unaffected. This effect finds its explanation in the mostly intra-layer character of magnons in FePS$_3$ with an inter-layer exchange coupling constant negligible compared to the intra-layer ones. As a result, we can tune the phonon energies with pressure at $B=0$, bringing a particular phonon (P$_3$) in resonance with the magnon and revealing the magnon-polaron resonance at $\textbf{k}=0$ involving a phonon and a doubly degenerate antiferromagnetic magnon. At resonance, the effect of the magnon-phonon interaction is maximized, lifting the magnon degeneracy and leading to three distinct coupled modes. This effect confirms the magnon-phonon interaction as the origin of the zero-field magnon splitting observed in Ref.~\cite{Vaclavkova2021,zhang2021}. For pressures above $4$~GPa, the magnon energy is decreased by $2$~cm$^{-1}$ as a result of the suppression of the interlayer antiferromagnetic interaction.

\subsection{Magneto-Raman scattering of magnon-polarons under high pressure}

\begin{figure*}[ht!]
\centering
\includegraphics[width=0.95\linewidth,angle=0,clip]{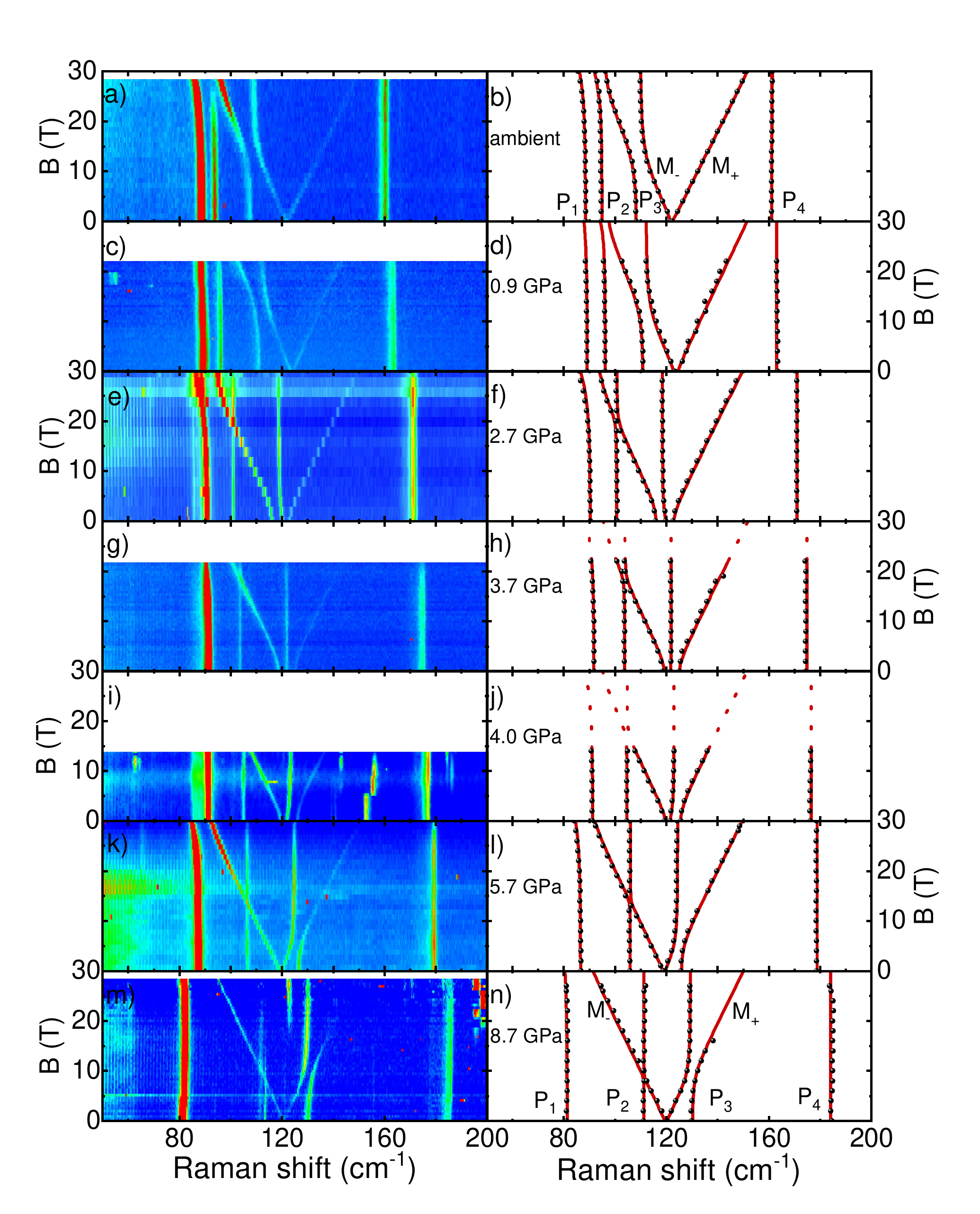}
\caption{(left) False color maps of the magneto-Raman scattering response of bulk FePS$_3$ measured at $T=4.5$~K and (right) evolution of the maxima of the peaks of the Raman scattering response (black dots) and of the calculated mode energies (red solid and dotted lines) as a function of the magnetic field for a,b) ambient pressure (from Ref.~\cite{Vaclavkova2021}), c,d) $P=0.89$~GPa, e,f) $P=2.7$~GPa, g,h) $P=3.7$~GPa, i,j) $P=4.0$~GPa, k,l) $P=5.7$~GPa and m,n) $P=8.7$~GPa.}
\label{Fig5}
\end{figure*}

Applying an external magnetic field on bulk FePS$_3$ lifts the antiferromagnetic magnon double degeneracy by adding a Zeeman contribution to the magnon energy $E_{M_{\pm}}=E^0_M \pm E_Z$ with $E_Z=g \mu_B B$, where $g$ is the Landé factor, $\mu_B$ is the Bohr magneton and $B$ the external magnetic field. It allows identifying the magnon excitation through its energy splitting into M$_-$ and M$_+$ branches, together with monitoring the energy detuning between phonons and the magnon branches. At ambient pressure, recent studies~\cite{McCreary2020,Liu2021,Vaclavkova2021,zhang2021} have reported a splitting of the magnon in FePS$_3$ in a magnetic field with a typical Landé factor of $g\sim2.1$. For magnetic fields up to $30$~T both branches of the magnon excitation can be shifted by $\pm 3.7$~meV ($\sim 30$~cm$^{-1}$) which allows tuning the M$_-$ magnon in resonance with the three lower-in-energy phonons P$_{1-3}$ and determining, at resonance, the three coupling constants $\lambda_{1-3}$ describing the interaction with phonons P$_{1-3}$, respectively.

The energy separation between the magnon and the higher-in-energy P$_4$ phonon ($40$~cm$^{-1}$) is too large to reveal this coupling with available magnetic fields and all magnon polarons in FePS$_3$ at ambient pressure involve the M$_-$ magnon branch. As we have seen in the preceding section, applying hydrostatic pressure to bulk FePS$_3$ allows changing the phonon spectrum with respect to the magnon excitation. In the investigated range of pressure, P$_3$ evolves from a situation at ambient pressure where its energy is lower than the magnon energy, to an energy resonance with the magnon close to $P=4$~GPa, and for higher pressures, the P$_3$ phonon energy is larger than that of the magnon. Bulk FePS$_3$ under pressure, with its unusually large magnon energy, represents a unique platform to investigate magnon-polarons in a large variety of magnon-phonon energy detuning, including the resonance.

The results of low-temperature magneto-Raman scattering experiments are presented in Fig.~\ref{Fig5} in the form of false color plots of the Raman scattering intensity as a function of magnetic field for different values of the pressure, together with the evolution of the maxima of the peaks of the Raman scattering response with the results of our calculation (solid and dashed lines). For all the investigated values of pressure, the measured $g$ factor is $g=2.1\pm0.15$ and does not change with pressure. At ambient pressure, we reproduce in Fig.~\ref{Fig5}a,b the experimental results from Ref.~\cite{Vaclavkova2021} showing the coupling of the M$_-$ magnon branch with P$_{1,2,3}$ evidenced by the avoided crossing between the modes at resonance. Increasing the hydrostatic pressure modifies the phonon spectrum as described in the first section, varying the energy detuning between P$_{1,2,3}$ and the magnon. Our data at P$=0.9$~GPa up to $B=22$~T, similarly to the ambient pressure case, show a large avoided crossing between P$_3$ and the M$_-$ branch~\cite{Liu2021,Vaclavkova2021}.

At higher pressure values, P$_3$ is brought in resonance with the magnon M and the Raman scattering response now displays three features of coupled magnon-phonon modes at $B=0$ in the magnon range of energy. This situation is presented in Fig.~\ref{Fig5}e-j. Applying a magnetic field in this case is of particular interest as it allows the identification of the magnon branches. For these three pressure values, the three-peaks structure observed at $B=0$ evolve in a very similar way: the higher and lower energy peaks disperse linearly with the magnetic field and they acquire a magnon-like character while the central peak sees its energy slightly changing in the first few Tesla (decreasing at $P=2.7$~GPa, staying rather constant for $P=3.7$~GPa and slightly increasing for $P=4.0$~GPa), and then becomes magnetic field independent with and energy position that matches the bare phonon energy. This behavior for $B<6$~T is shown in more details in Fig.~S4 of the supplementary materials, and we present in Fig.~S5 polarization resolved Raman scattering spectra measured at $P=3.7$~GPa and at $B=0$ showing that the magnon features are only observed in a co-linear configuration while the central peak, observed in both co- and cross-linear polarization configuration, shows a phonon-like behavior.

These evolutions are a direct signature of the magnon-phonon interaction: from a situation where the magnon-phonon interaction is resonant because of the coincidence of the bare magnon and P$_3$ phonon energies, leading to the observation of three strongly coupled magnon-phonon modes, to the situation in which the magnetic field disentangles these three coupled modes by inducing an energy detuning between the magnon branches and P$_3$. As a result, the interaction is reduced and for high enough magnetic fields for which the two magnon branches M$_-$/M$_+$ are tuned away from P$_3$, the central peak recovers mainly a phonon-like character and its energy matches the bare phonon energy.

At $P=2.8$~GPa, for magnetic fields close to $B=20$~T where the M$_-$ branch is tuned to P$_2$, the Raman scattering response shows a simple crossing between these two excitations, with no apparent sign of interaction. When pressure is applied, the coupling of the M$_-$ component of the magnon with the P$_2$ phonon vanishes and becomes hardly observable. This absence of coupling is not specific to this particular value of pressure and can also be observed at higher pressures, e.g. at $P=3.7-5.7-8.7$~GPa. The coupling between M$_-$ and P$_1$, on the other hand, remains and can be observed at the highest magnetic fields.

Increasing further the pressure (above $P=4$~GPa), the P$_3$ phonon has an energy larger than that of the magnon. The energy of the M$_+$ component increases with the magnetic field and can tuned to the P$_3$ phonon energy. At resonance, these two excitations show an avoided crossing behavior typical of strongly coupled modes, but with a coupling strength that appears smaller than the one at ambient pressure. The M$_-$ component shifts down in energy with increasing magnetic field and crosses the P$_2$ phonon mode with no sign of interaction. For magnetic fields above $B=25$~T, the interaction with the P$_1$ phonon is observed, even though at $P=8.7$~GPa when FePS$_3$ is in its HP-I phase, the P$_1$ phonon energy is strongly reduced and the magnon-$P_1$ energy detuning stays large in the investigated range of magnetic fields. 

\begin{figure}
\centering
\includegraphics[width=0.95\linewidth,angle=0,clip]{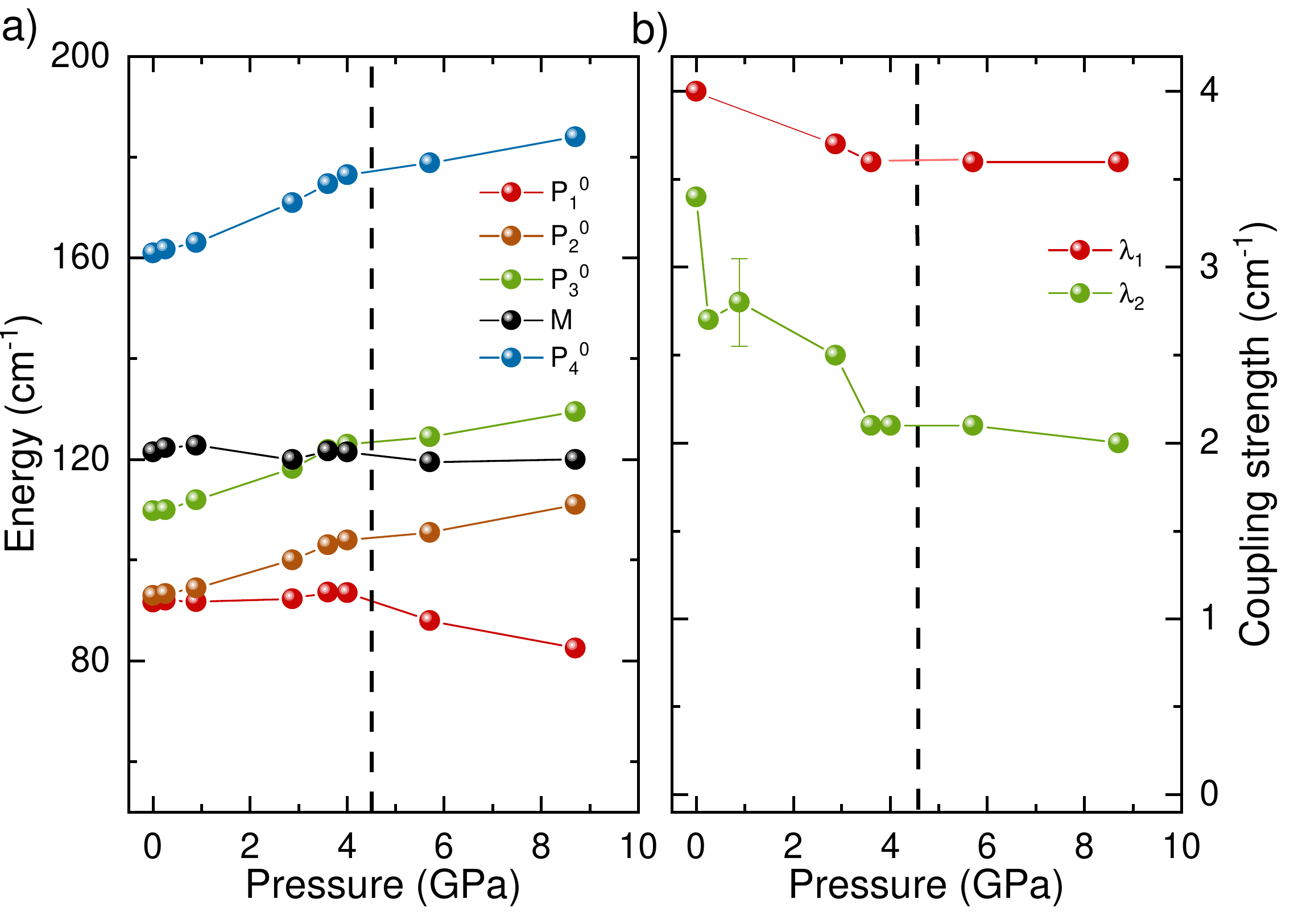}
\caption{a) Evolution of the bare phonon and magnon excitation energies used in our simulations as a function of the applied pressure. b) Evolution of the coupling constants for the different phonon modes as a function of the applied pressure. We have indicated in the figure a typical error bar on the coupling constant.
\label{Fig6}}
\end{figure}

To describe our magneto-Raman scattering data, we have used the phenomenological model of coupled magnon-phonon modes described by the Hamiltonian \ref{Ham}. We take into account the fact the coupling to P$_2$, comparable to the other coupling constants at ambient pressure, vanishes and is absent from the data at $2.7$~GPa and for higher pressure values. This is puzzling as no phase transitions are expected up to this value of hydrostatic pressure and hence, the phonon symmetry and its displacement pattern are unchanged with respect to the ones at ambient pressure. We are then led to introduce a pressure independent coupling between P$_1$ and P$_2$ phonons, and decoupling P$_2$ from the magnon. Within this frame, P$_2$ inherits the properties of P$_1$, in particular its potential coupling to the magnon, if the P$_1$-P$_2$ energy detuning is small compared to the coupling parameter $\beta$. This model describes well the experimentally observed behavior using $\beta=3$~cm$^{-1}$ but we can hardly define the nature of the interaction between the two phonons and this effect should be described on theoretical grounds. From these experiments and from the modelling, we deduce the pressure evolution of the different magnon-phonon interaction parameters together with the bare modes energies, see Fig.~\ref{Fig6}a,b. The magnon-polaron resonance involving P$_1$, clearly observed in our experiments performed at $P=2.7$~GPa and at $P=5.7$~GPa, does not dependent, within the error bar, on the applied hydrostatic pressure, even in the high pressure phase HP-I above $P=4.5$~GPa at which a coupling constant very close to that at ambient pressure is measured. At $P=8.7$~GPa, the energy of the P$_1$ phonon is lower than that at ambient pressure and the resonance cannot be reached but a change of the P$_1$ phonon energy at the highest magnetic fields allows the evaluation of the coupling constant. This softening of the P$_1$ mode can be related to high pressure metallic phase as it has been observed in different systems close to a metallic transition~\cite{Bellin2020}. The coupling constant with the P$_3$ phonon decreases with pressure from $\lambda_3=3.1\pm0.2$~cm$^{-1}$ at ambient pressure down to $\lambda_3=1.8\pm0.2$~cm$^{-1}$ at $P=8.7$~GPa. The deduced value for the $\lambda$ parameters are small with respect to the modes energies and can be considered as small perturbations, but are comparable to the full width at half maximum which allows for the proper observation of the avoided crossing at resonance.

\section{Conclusions}

To conclude, applying high pressure to bulk FePS$_3$ leads to a modification of its phonon spectrum but weakly affects magnons, the energy of which is defined mainly by intra-layer exchange parameters. Phonons and magnons are intrinsically coupled in bulk FePS$_3$ and we have shown that a particular phonon (P$_3$) can be tuned through the doubly degenerate antiferromagnetic magnon excitation. This leads to the formation of $\mathbf{k}=0$ magnon-polarons involving a twice degenerate magnon mode. They are evidenced in our low temperature micro-Raman scattering experiments under high pressure by a particular avoided crossing behavior involving the three modes when the resonance is achieved. Tuning pressure allows changing the hybridization between the magnon and the P$_3$ phonon in bulk FePS$_3$ in its magnetically ordered state. When the P$_3$ phonon mode is tuned in resonance with the magnon mode, the two components of the antiferromagnetic magnon excitation are splitted by $7$~cm$^{-1}$ by the magneto-elastic interaction and three modes clearly appear in the Raman scattering response. An external magnetic field facilitates the identification of the different coupled modes by monitoring the energy detuning between the magnons and the phonons. For pressures above $4$~GPa, we observe magnon-polarons involving the M$_+$ component of the magnon and the P$_3$ phonon. We also observe that the magneto-elastic interaction with the P$_2$ phonon vanishes when pressure is applied. We propose that there is coupling, yet to be defined, between P$_1$ and P$_2$ phonons of the magnetically ordered phase which makes the observation of magnon-polarons involving P$_2$ possible only for small P$_1$-P$_2$ energy detuning. Bulk FePS$_3$ with its high energy magnon excitation represents an ideal system to investigate magnon-phonon interaction in a rich variety of situations of phonons and magnons energy detuning.

\begin{acknowledgements}

We acknowledge technical support from I. Breslavetz. This work has been partially supported by the ANR projects ANR-17-CE24-0030 and ANR-19-CE09-0026 and by the EC Graphene Flagship project. M-A. M. acknowledges the support from the ERC (H2020) (Grant agreement No. 865826). We acknowledge the support of the LNCMI-EMFL, CNRS, Univ. Grenoble Alpes, INSA-T, UPS, Grenoble, France. D. R. and M. C. acknowledge  support  from  the  ANR project  ACCEPT  (Grant  No. ANR-19-CE24-0028). This  work  was  granted  access  to  the  HPC  resources  of IDRIS,  CINES  and  TGCC  under  the  allocation  2021-A0100912417 made by GENCI.

\end{acknowledgements}

%

\pagebreak

\widetext

\begin{center}
\textbf{\large Supplemental Materials: High pressure tuning of the magnon-polaron resonance in the layered antiferromagnet FePS$_{3}$}
\end{center}

\setcounter{equation}{0}
\setcounter{figure}{0}
\setcounter{table}{0}
\setcounter{page}{1}
\makeatletter
\renewcommand{\theequation}{S\arabic{equation}}
\renewcommand{\thefigure}{S\arabic{figure}}
\renewcommand{\thetable}{S\arabic{table}}
\renewcommand{\bibnumfmt}[1]{[S#1]}
\renewcommand{\citenumfont}[1]{S#1}

In this supplementary materials, we provide additional experimental data supporting our main findings presented in the manuscript and a comparison of the model used in the main text with a simpler Hamiltonian based on a $3\times3$ matrix.

\subsection{Evolution of high energy phonons with pressure}

\begin{figure}[h]
\includegraphics[width=0.7\linewidth,angle=0,clip]{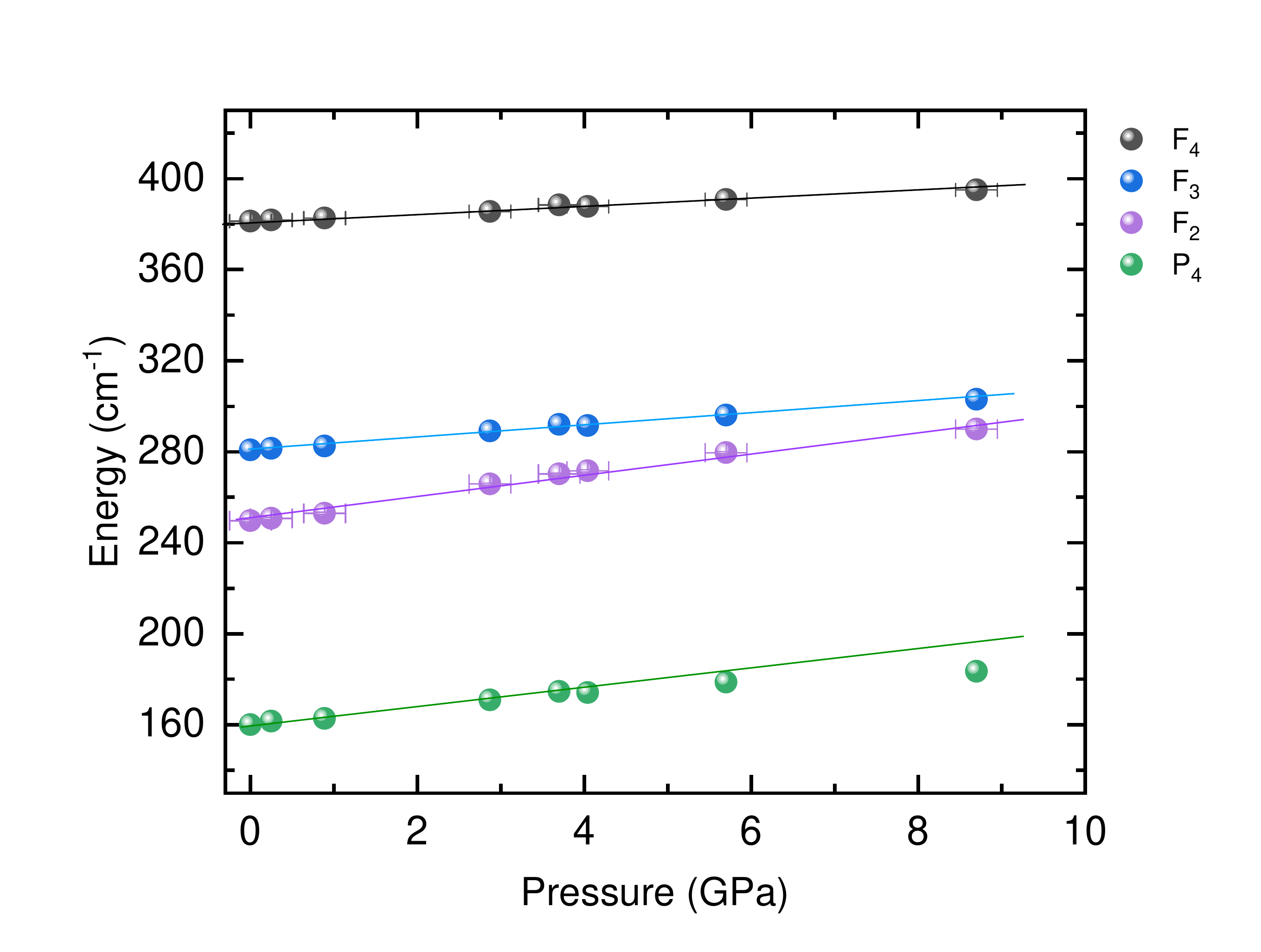}
\caption{Evolution at $T=4.5$~K of the phonon energies as a function of the applied pressure. Solid lines are guides for the eye.
\label{FigHE}}
\end{figure}

Fig.~\ref{FigHE} shows the evolution of the energies of the high energy phonons in bulk FePS$_3$ as a function of pressure. In contrast the phonon modes below $160$~cm$^{-1}$, the high energy phonons evolve linearly with pressure with little changes at the structural phase transition (P$_C=4.6$~GPa). Table~\ref{tableS1} indicates the phonon and magnon energies at ambient pressure together with their increasing rates with pressure obtained by a linear fit for $P < 3$~GPa.

\begin{table}[h]
\begin{tabular}{|l|c|c|}
  \hline
  Label & Energy (cm$^{-1}$)  & R (cm$^{-1}$/GPa) \\
  \hline
  P$_1$ & 88.5 & 0.95 \\
  P$_2$ & 93.7 & 2.7 \\
  P$_3$ &  107.3   & 3.0    \\
  M &  121   & 0 \\
  P$_4$ & 160.5   & 3.8 \\
  F$_1$ & 230.8   &  \\
  F$_2$ & 249 &  5.1\\
  F$_3a$ & 280.4    & 2.8 \\
  F$_3b$ & 282.6    & 3.1 \\
  F$_4$ & 381.2    & 1.7 \\
  \hline

\end{tabular}
\caption{Experimentally measured energies at ambient pressure and at $T=4.5$~K of the different peaks of the Raman scattering spectrum together with their evolution rates when applying pressure.}
\label{tableS1}
\end{table}

\clearpage

\subsection{Calculations of phonon energies as a function of pressure}

In Tab.~\ref{table:1} we report the evolution of the four lowest vibrational modes as a function of pressure computed via density functional theory (DFT) and the PBE0 exchange correlation functional. We have performed calculations up to 2 GPa, after that the accuracy of the Gaussian basis set was found not be reliable as atoms starts getting too close to each other. From our calculations, we can deduce that the phonon modes evolve when applying pressure with rates of $0.5$, $2.65$ and $2.275$~cm$^{-1}$/GPa for P$_1$ ($B^1_g$), P$_2$ ($B^2_g$) and P$_3$ ($B^3_g$), respectively.  Moreover, because no structural phase transition occurs in the analyzed range of pressure, the patterns of the normal modes normalized to classical amplitudes do not evolve between ambient pressure and P=$2$~GPa

\begin{table}[h!]
    \centering
\begin{tabular}{ c | c | c | c | c | c}
\hline
\hline
Pressure (GPa) & P$_1$ - $B^1_g$  (cm$^{-1}$) & P$_2$  - $B^2_g$  (cm$^{-1}$) & P$_3$  - $A^1_g$  (cm$^{-1}$) & P$_3$  - $B^3_g$  (cm$^{-1}$) & P$_4$  - $A^2_g$   (cm$^{-1}$) \\
\hline
0 & 94.63 & 94.70 & 115.44 & 125.36 & 159.67\\
1 & 95.71 & 97.80 & 115.45 & 128.38 & 161.28\\
2 & 95.71 & 99.94 & 115.56 & 129.91 & 164.56\\
\hline
\hline
\end{tabular}
\caption{Theoretical values for the four lowest vibrational modes as a function of pressure computed via DFT and PBE0 exchange-correlation functional. $A_g$/$B_g$ represents the symmetry of the vibrational mode as labelled in the main text.}
\label{table:1}
\end{table}

\clearpage

\subsection{Comparison with a simpler $3\times3$ Hamiltonian}

We consider the following Hamiltonian to describe the coupling between three modes, two modes being degenerate and non interacting with each other, and a third mode (phonon) interacting with the two degenerated modes:

\begin{equation}
H = \begin{bmatrix}
M_+ & 0 & \delta \\
0 & M_- & \delta \\
\delta & \delta & P
\end{bmatrix}
\end{equation}

where $M_{\pm}=E_M \pm g \mu_B B$ are the two magnon components, $\delta=2.6$~cm$^{-1}$ is the magnon-phonon interaction parameter, P is the phonon energy. The magnon-phonon energy detuning ($\Delta=E_M-E_{ph}$) can be changed by applying pressure.

\begin{figure}[h]
\includegraphics[width=0.7\linewidth,angle=0,clip]{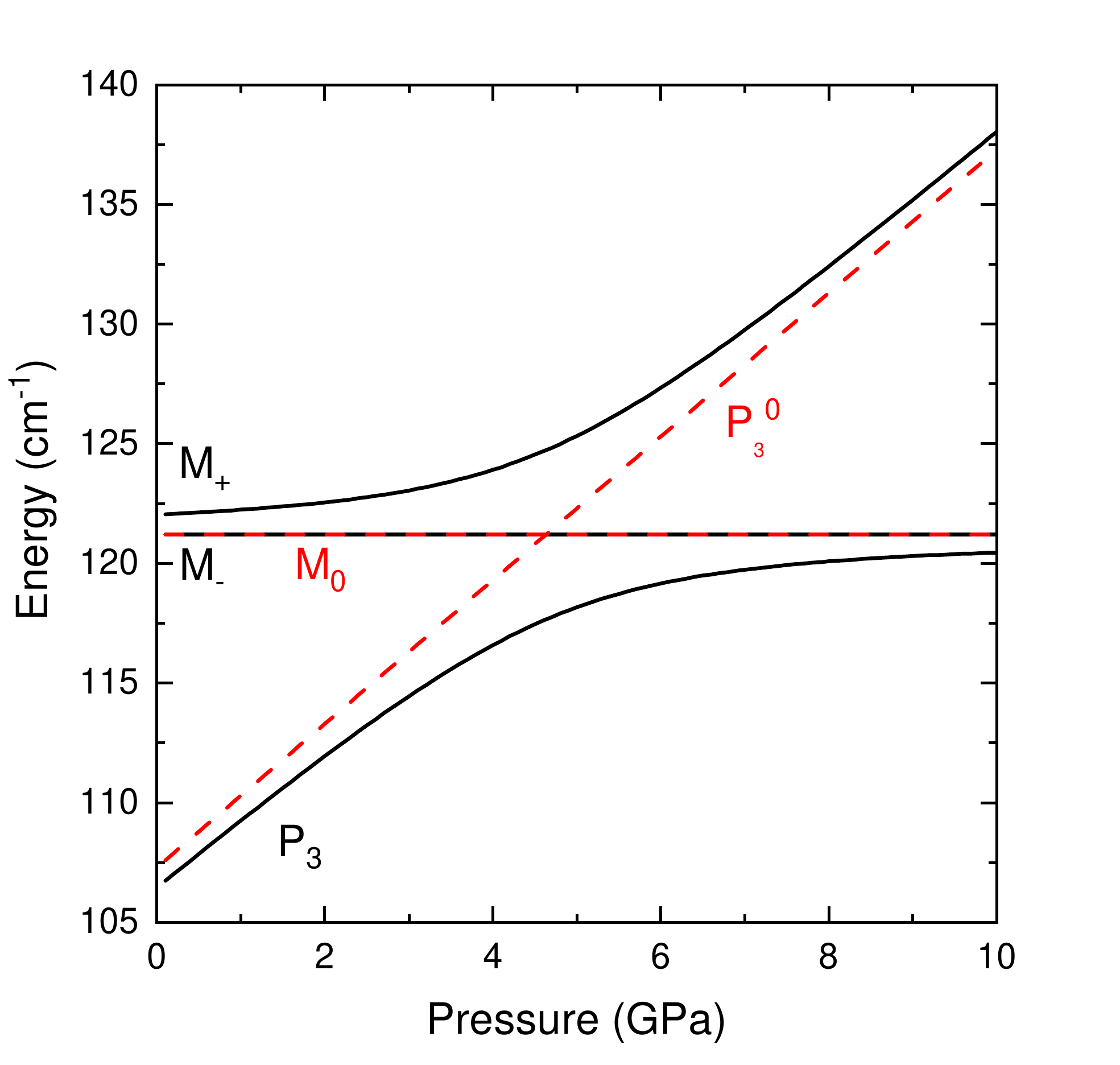}
\caption{Evolution of the eigenvalues of the $3 \times 3$ matrix as a function of the hydrostatic pressure (phonon energy). The magnon energy is set at $E_M=121$~cm$^{-1}$. The red dashed line indicate the bare magnon and phonon energies used in the calculation.
\label{Ham3x3P}}
\end{figure}

\begin{figure}[h]
\includegraphics[width=0.7\linewidth,angle=0,clip]{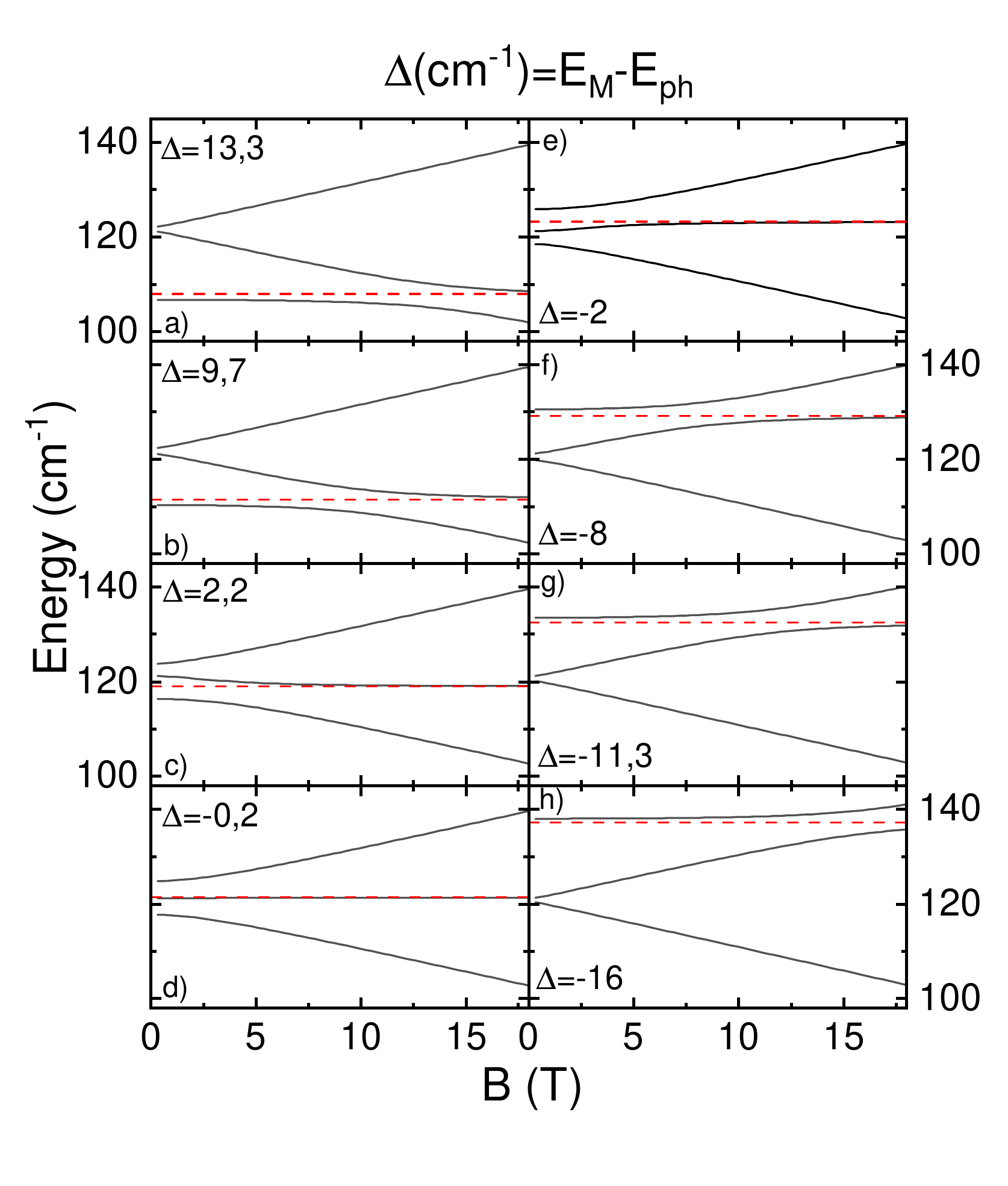}
\caption{Evolution of the eigenvalues of the $3 \times 3$ matrix as a function of the magnetic field for different magnon-phonon energy detuning ($\Delta$). The magnon energy is set at $E_M=121.2$~cm$^{-1}$ and $g=2.1$. The red dashed line indicate the bare phonon energy used in each panel.
\label{modHam}}
\end{figure}

Fig.~\ref{Ham3x3P} shows the evolution of the eigenvalues of the $3 \times 3$ Hamiltonian as a function of the applied pressure at $B=0$~T. The magnon-phonon interaction lifts the magnon degeneracy at any value of the magnon-phonon energy detuning. At resonance, one the initially magnon mode shows an avoided crossing behavior with the phonon mode while the other magnon component stays at the same energy. Its character on the other hand changes from magnon-like to phonon-like at resonance, back to magnon-like at high pressure.

\begin{figure}[ht]
\includegraphics[width=0.7\linewidth,angle=0,clip]{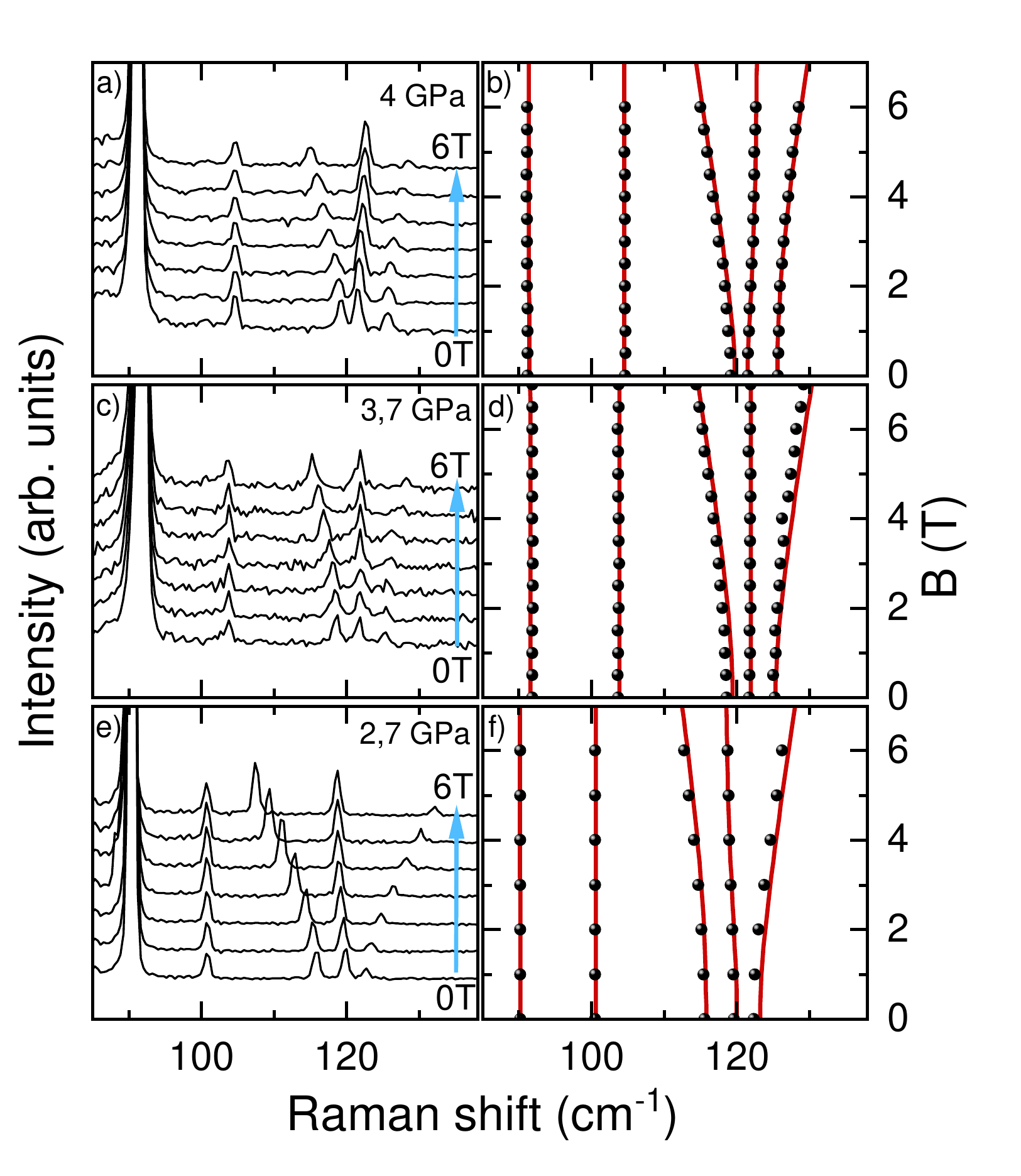}
\caption{Raman scattering spectra at different values of the magnetic field up to $B=6$~T and results of our modelling (red lines) using the model described in the text at $P=4.0$~GPa (a,b), $P=3.7$~GPa (c,d) an $P=2.7$~GPa (e,f), illustrating the three situations for which the phonon energy is above (a,b), quasi degenerate with (c,d), and below (e,f) the magnon energy.
\label{FigLowB}}
\end{figure}

The strong similarity between the evolutions calculated using the $3\times3$ Hamiltonian (see Fig.~\ref{modHam}) and the experimental results together with the results of the full model (see Fig.\ref{FigLowB}) indicate that most of the physics of this system is grasped by the $3\times3$ Hamiltonian, including the non linear evolutions close to the magnon-phonon resonance. This simpler model describes satisfactorily the magnon-phonon interaction when a two-fold degenerated magnon mode is involved. In particular, it describes precisely the behavior close to the resonant magnon-phonon condition (detuning $\Delta \sim 0$). At $B=0$, when the enegy detuning is close to zero, the magnon phonon interaction lifts the magnon degeneracy and within a significant range of energy detuning, the phonon mode lies in between the two magnon excitation (see Fig.~\ref{modHam}c,d,e).

Fig.~\ref{FigLowB} shows the low magnetic field experimental Raman scattering spectra together with their analysis using the full model developed in the main text, for $P=2.7 - 3.7$ and $4$~GPa.

\clearpage

\subsection{Linear polarization resolved Raman scattering response}

\begin{figure}[ht]
\includegraphics[width=0.7\linewidth,angle=0,clip]{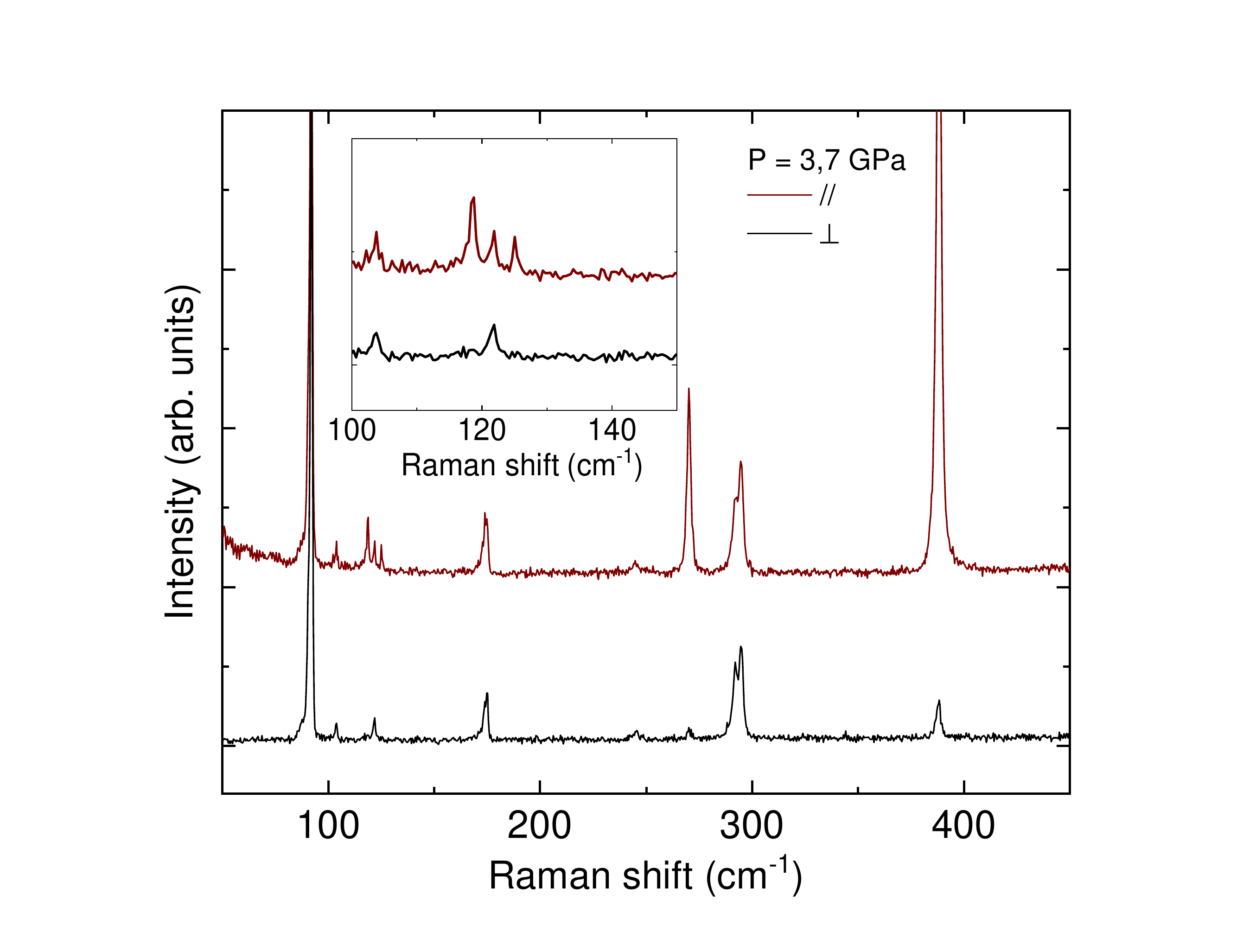}
\caption{Linear-polarization resolved low temperature Raman scattering spectrum of bulk FePS$_3$ at $P=3.7$~GPa. The black (brown) line is the crossed- (co-) linear polarization configuration. Inset: Zoom on the magnon range of energy.
\label{Figcocross}}
\end{figure}

Fig.~\ref{Figcocross} presents the linearly polarized Raman scattering response of bulk FePS$_3$ using the same linear polarization for excitation and collection (colinear polarization) of orthogonal linear polarizations (cross linear polarization). The two "magnon-like" peaks are only visible in the colinear polarization while the central "phonon-like" peak appears in both co- and cross- linear polarizations.

\end{document}